\documentclass[fleqn,12pt,twoside]{article}
\usepackage{espcrc1}

\usepackage{graphicx}
\usepackage[figuresright]{rotating}

\newcommand{\AmS}{{\protect\the\textfont2
  A\kern-.1667em\lower.5ex\hbox{M}\kern-.125emS}}

\title{$\rm ^{26}Al$ and $\rm ^{60}Fe$ from massive stars}

\author{M. Limongi\address{INAF - Osservatorio Astronomico di Roma, \\ 
        Via Frascati 33, I-00040 Monteporzio Catone, Roma, Italy} 
        and
        A. Chieffi\address{Istituto di Astrofisica Spaziale e Fisica Cosmica (CNR), \\
        Via Fosso del Cavaliere, I-00133, Roma, Italy}
        }
       
\begin{document}

\maketitle

\begin{abstract}
We discuss the production of $\rm ^{26}Al$ and $\rm ^{60}Fe$
in two massive star models, namely a 25 and a 80 $\rm M_\odot$, of initial 
solar metallicity. 
\end{abstract}

\section{Introduction} 
The radioactive $\rm ^{26}Al$ and $\rm ^{60}Fe$ isotopes 
decay into their daughter stable nuclei with lifetimes of the order of a few $\sim 10^{5}$ 
yr. These times are short enough, compared to the typical timescales of Galaxy 
evolution, that the observations of these isotopes clearly identify the locus of the present 
nuclesynthetic enrichment
through the Galaxy. Observations of the $\rm \gamma-ray$ emission from the 
radioactive $\rm ^{26}Al$ at 1.809 MeV have been reported by several missions 
over the last 15 years and show that $\rm ^{26}Al$ is mostly confined in the disk of 
the Galaxy with some prominent areas of emission corresponding to the Vela and 
Cygnus regions (Diehl \& Timmes 1998).
Many sources can be responsible for the production of $\rm ^{26}Al$
in the Galaxy, among them Core Collapse Supernovae, WR stars, Novae and AGB stars.
Unfortunately the observations do not provide a strong constraint on 
the dominant source of production yet. There is another measurable
quantity, however, that could help to distinguish between the various
competing sources. In particular $\rm \gamma-rays$ from the radioactive
$\rm ^{60}Fe$ may be a good discriminant for the origin of Galactic
$\rm ^{26}Al$. Indeed, core collapse supernovae should produce comparable
amounts of both $\rm ^{26}Al$ and $\rm ^{60}Fe$ while all the other candidates
produce a much smaller amount of $\rm ^{60}Fe$ compared to $\rm ^{26}Al$.

A few years ago we started a series of papers devoted to the study of the 
presupernova evolution and explosive nucleosynthesis of massive stars, in the 
range 13-35 $\rm M_\odot$, as a function of initial mass and metallicity (Chieffi 
et al. 1998, Limongi et al. 2000, Chieffi \& Limongi 2004). No mass loss
was included into those computations. At present we are addressing the 
presupernova evolution and explosion of massive stars with mass loss
in a more extended range of mass, namely between 13 and 120 $\rm M_\odot$
(Chieffi \& Limongi in preparation).
In this paper we preliminarily discuss the production of $\rm ^{26}Al$ and $\rm 
^{60}Fe$ in two selected massive star models, namely a 25 and a 80 $\rm M_\odot$, of 
initial solar metallicity. 

\begin{figure}[ht]
\begin{center}
\includegraphics[scale=0.45]{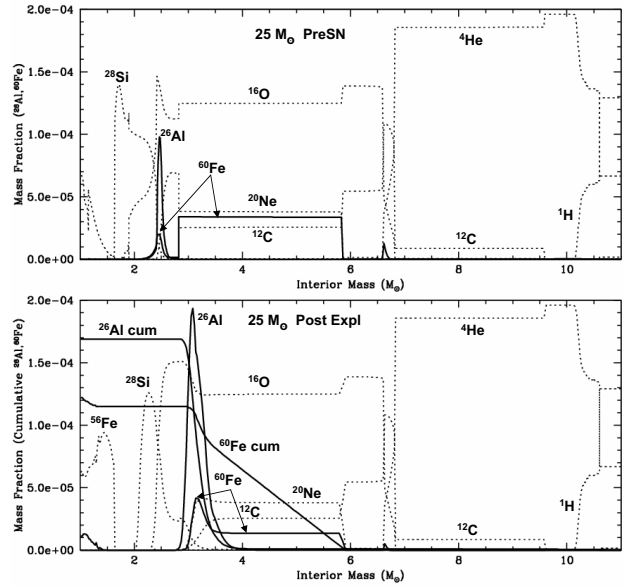}
\end{center}
\caption{{\em Upper Panel}: internal profile of $\rm ^{26}Al$ and $\rm ^{60}Fe$ (thick solid lines) 
for the 25 $\rm M_\odot$ model
before the explosion. The dotted lines refer to the most
abundant isotopes. The y-axis of these lines is not shown and ranges between 0 and 1.
{\em Lower Panel}: internal profile of $\rm ^{26}Al$, $\rm ^{60}Fe$ together with their
cumulative values (thick solid lines) for the 25 $\rm M_\odot$ model after the explosion.
Also in this case the most abundant isotopes are shown. The y-axis of $\rm ^{26}Al$ and 
$\rm ^{60}Fe$ is not shown and ranges between 0 and $5\cdot 10^{-4}$.}
\end{figure}

\section{Production of $\rm ^{26}Al$ and $\rm ^{60}Fe$ in massive stars}
In stars, $\rm ^{26}Al$ is produced by proton capture on $\rm ^{25}Mg$
and destroyed by $\rm \beta^{+}$ decay into $\rm ^{26}Mg$ and by $(n,p)$,
$(n,\alpha)$ and $(p,\gamma)$ reactions (Clayton \& Leising 1987). 
In massive stars ($\rm M\ge 11~M_\odot$) $\rm ^{26}Al$
can be produced by (1) hydrostatic H burning, (2) hydrostatic C and Ne burning and
(3) explosive C and Ne burning. Any $\rm ^{26}Al$ produced by these stars
is ejected by both stellar wind and explosion in different proportions
depending on the initial mass.

$\rm ^{60}Fe$ is produced by the sequence $\rm ^{58}Fe(n,\gamma)^{59}Fe(n,\gamma)^{60}Fe$. 
This sequence is mainly powered by 
the $\rm ^{22}Ne(\alpha,n)^{25}Mg$ reaction, that is efficient during (1) 
hydrostatic He burning, (2) hydrostatic C burning and (3) explosive Ne burning.

In this paper we discuss the production of $\rm ^{26}Al$ and $\rm ^{60}Fe$ in
two massive star models, namely a 25 and a 80 $\rm M_\odot$, of 
initial solar metallicity. These models have been computed with the latest
version of the FRANEC code (Chieffi \& Limongi 2004). Mass loss have been included
following the prescriptions of (1) Vink et al. (2000, 2001) for the blue supergiant
phase ($\rm T_{eff}>12000~K$), (2) de Jager et al. (1988) for the red supergiant
phase ($\rm T_{eff}<12000~K$) and (3) Nugis \& Lamers (2000) during the Wolf-Rayet phase.
The explosion has been computed adopting a PPM hydro code in the framework of the piston method.
The initial velocity of the piston is set in order to eject all the matter above the iron core.
As a consequence the final kinetic energy is different depending on the progenitor mass.

Figures 1 and 2 show the internal profiles of both $\rm ^{26}Al$ and $\rm ^{60}Fe$
for the 25 and a 80 $\rm M_\odot$ models respectively. The upper panels refer to the
pre explosive abundances while the lower panels refer to the final explosive ones.

\begin{figure}[ht]
\begin{center}
\includegraphics[scale=0.45]{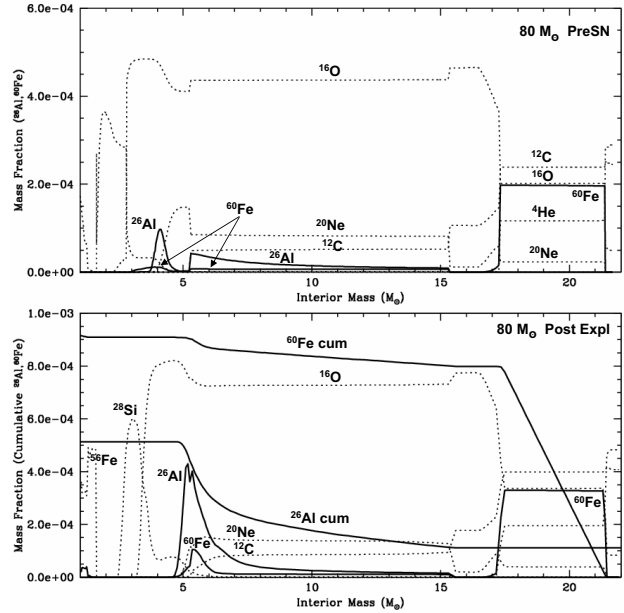}
\end{center}
\caption{Same as Figure 1 but for the 80 $\rm M_\odot$ model. The y-axis of $\rm ^{26}Al$ and 
$\rm ^{60}Fe$ in the lower panel is not shown and ranges between 0 and $6\cdot 10^{-4}$.}
\end{figure}

In the 25 $\rm M_\odot$ star the bulk of $\rm ^{26}Al$ is produced by the Ne shell
during the presupernova evolution; then it is completely destroyed by the explosion and
reproduced, farther out in mass, by the explosive Ne and C burning. 
Hence, in this case, the final yield of $\rm ^{26}Al$ ($\rm 1.69\cdot 10^{-4}~M_\odot$)
is dominated by the explosive nucleosynthesis. On the contrary the $\rm ^{60}Fe$ produced
by the C convective shell during the presupernova evolution is only marginally
affected by the explosive Ne and C burning. Hence the total final yield of $\rm ^{60}Fe$
($\rm 1.15\cdot 10^{-4}~M_\odot$) is due, by more than 70\%, to its hydrostatic production. 
Let us eventually note that no $\rm ^{60}Fe$ is synthesized in a sizeable amount 
within the He convective shell.

In the 80 $\rm M_\odot$ star the production of $\rm ^{26}Al$ is more complicated than 
in the 25 $\rm M_\odot$. Indeed $\rm ^{26}Al$ is initially produced during the 
presupernova evolution by the central H burning. Roughly half of this $\rm 
^{26}Al$ is lost by stellar winds during the Wolf-Rayet phase, while the 
remaining is burnt during the following He burning. Another episode of $\rm 
^{26}Al$ production occurs during the subsequent C and Ne shell burning phases 
so that the final profile of $\rm ^{26}Al$, that is present within the star at 
the moment of the explosion, is shown in the upper panel of Figure 2. The effect 
of the explosion is to completely destroy the $\rm ^{26}Al$ produced by the 
hydrostatic Ne burning shell and to reproduce it farther out in mass. The final 
total yield of $\rm ^{26}Al$ ($\rm 5.13\cdot 10^{-4}~M_\odot$) is due to the sum 
of the $\rm ^{26}Al$ present in the stellar wind ($\rm 1.11\cdot 10^{-
4}~M_\odot$), the one that survived the explosive nucleosynthesis ($\rm 1.29\cdot 
10^{-4}~M_\odot$) and the one due to the explosive burning ($\rm 2.73\cdot 
10^{-4}~M_\odot$). At variance with the 25 $\rm M_\odot$ model, the He 
convective shell dominates the final nucleosynthesis of $\rm ^{60}Fe$ (see the 
upper panel of Figure 2). Such an occurrence is due to the fact that, in this 
case, the He convective shell is exposed to a temperature high enough
that the $\rm ^{22}Ne(\alpha,n)^{25}Mg$ reaction becomes extremely efficient. 
Such a high temperature obviously induces also a large He consumption ($\sim 
0.195$ in mass fraction) within the He convective shell. 
The final total yield of $\rm ^{60}Fe$ 
($\rm 9.09\cdot 10^{-4}~M_\odot$) is due to the sum of the $\rm ^{60}Fe$ 
produced by the He convective shell ($\rm 7.99\cdot 10^{-4}~M_\odot$), the one 
produced by the C convective shell ($\rm 6.4\cdot 10^{-5}~M_\odot$) and the one 
produced by explosive Ne and C burning ($\rm 4.6\cdot 10^{-5}~M_\odot$). 
No $\rm ^{60}Fe$ is present within the stellar wind.

\section{Summary and conclusions}
We have discussed the production of $\rm ^{26}Al$ and $\rm ^{60}Fe$ in two massive
star models, namely a 25 and a 80 $\rm M_\odot$, of initial 
solar metallicity. We have shown that in the 25 $\rm M_\odot$ model,
the final yield of $\rm ^{26}Al$ is dominated by 
explosive C and Ne burning while $\rm ^{60}Fe$ 
mainly comes from the convective C shell ($\sim 70\%$)
and to a lesser extent ($\sim 30\%$) by explosive C and Ne burning.
On the contrary, in the 80 $\rm M_\odot$ model
$\rm ^{26}Al$ comes from: stellar
wind ($\sim 20\%$), convective C shell ($\sim 25\%$)
and explosive C and Ne burning ($\sim 55\%$).
The final yield of $\rm ^{60}Fe$ is dominated by the He convective shell
by more than 85\%. No $\rm ^{60}Fe$ is present in the stellar wind.
We explored the dependence of the final $\rm ^{26}Al$ yield on the explosion 
energy and we found that it remains essentially constant up to, at least, final 
kinetic energies of the order of a few foes.

As a final comment let us note that the $\rm ^{60}Fe/\rm ^{26}Al$ ratio is 0.68 
and 1.77 for the 25 and the 80 $\rm M_\odot$ models respectively. These ratios 
are both well above the upper limit ($\sim 0.20$) reported by the $\gamma$-ray 
instruments presently in Space. A detailed discussion of 
the theoretical steady state $\rm ^{60}Fe/\rm ^{26}Al$ prediction,
in the light of the present uncertainties in the computation
of the stellar models, will be given in a forthcoming paper.

\end{document}